\documentstyle[12pt,aps,floats,prc,epsfig,tighten]{revtex}
\def\Journal#1#2#3#4{{ #1} {\bf #2}, #3 (#4) }
\def\NPA{{ Nucl. Phys.} }
\def\PRL{Phys. Rev. Lett.}

\def\PRC{{Phys. Rev.} C}

\def\PL {Phys. Lett.}   
\newcommand{\matrixel}[3]{\mbox{$\left<#1|#2|#3\right>$}}                   
\begin{document}
\title{Interaction induced deformation in momentum distribution of
spin polarized nuclear matter}
\author{T. Frick,  H. M\"uther, A. Sedrakian}
\address{Institut f\"ur
Theoretische Physik,  Universit\"at T\"ubingen, D-72076 T\"ubingen, Germany}
\maketitle
 
\begin{abstract}
Effects of spin polarization on the structure of symmetric nuclear matter and
pure neutron matter are investigated. We show that the spin polarization
induces a deformation of the Fermi spheres for nucleons with spin
parallel and opposite to the polarization axes. This feature can be related to
the structure of the one pion exchange contribution to a realistic
nucleon-nucleon interaction. While the anisotropies in the momentum
distribution  lower  the energy of the system by small amount, the
associated variations of the single particle energies with the angle 
between the polarization axis and the particle momentum are significant.
\end{abstract}

The magnetic susceptibility of nuclear matter has received a considerable
amount of attention as this quantity is relevant for the studies in
astrophysics of compact objects. The strong magnetic fields in the interiors of
neutron stars may lead to a magnetization of the material of nuclear matter,
which implies a spin polarization of the interacting nucleons. Various groups
have  discussed  the possibility of a phase transition of normal neutron
matter to a ferromagnetic
state\cite{silv,rice,ostg,clark,panda,fantoni,vidana}.  With realistic models
of the nucleon-nucleon (NN) interaction, however, such a phase transition seems
to be suppressed for neutron matter up to densities well above the saturation
density $\rho_0$ of symmetric nuclear matter.  Moreover, it  turns out that the
NN interaction even tends to reduce the susceptibility below the value of a
non-interacting Fermi gas
\begin{equation}
\chi_F = \frac{\mu^2 m k_F}{\hbar^2\pi^2}\,,
\end{equation}   
where $\mu$ and $m$ represent the magnetic moment and the mass of the
neutron while $k_F$ stands for the Fermi momentum of system. This reduction 
is related to the density of states at the Fermi surface, which can be expressed
in terms of an effective mass $m^*$, which is typically smaller than the bare
nucleon mass $m$, and the strength of the spin-spin interaction, represented by
a Landau parameter $G_0$. An estimate for the susceptibility is then
\begin{equation}
\frac{\chi}{\chi_F} \approx \frac{m^*}{m \left( 1 + G_0 \right)}\,.
\end{equation}
If, however, the spin-spin interaction becomes attractive at densities
above the nuclear saturation density, as it is the case for
some effective interactions like the Skyrme force, the system may undergo a
phase transition to a ferromagnetic state\cite{vidau,reddy}.

The polarizability of nuclear matter or its spin-spin correlation function is
also an important ingredient for the study of the interaction of neutrinos with
nuclear matter. This interaction of neutrinos with dense matter is of high
importance for the simulation of supernova explosion and the cooling of neutron
stars. In order to compute the mean free path of neutrinos in nuclear matter
one has to determine the nuclear response function for density and spin
excitations\cite{reddy,sawyer,navarro,dalen} in particular for lower energies.
These response functions are essentially characterized by the single-particle
spectrum of the nucleons with momenta $p$ around the Fermi momentum, which is
often  parameterized in terms of an effective mass
and momentum independent single-particle potential $U$
\begin{equation}
\varepsilon(p) = \frac{p^2}{2m^*} + U\,, \label{eq:epsk}
\end{equation}  
and the residual interaction, which is conveniently characterized by the Landau
parameters $F_0$ and $G_0$ for the density and spin response functions,
respectively.

The studies of the nuclear polarizability and the
nuclear response functions all consider a conventional 
model for nuclear matter,
where the single-particle states are occupied
according to the magnitude of the single-particle momentum $p$. As a typical
example we refer to the Brueckner-Hartree-Fock (BHF) calculation of Vida\~na et
al.\cite{vidana}, who evaluate the single-particle energies for nucleons with
momentum $p$ and spin projection $s$ according to
\begin{equation}
\varepsilon (p,s) =  \frac{p^2}{2m} + \sum_{s'} 
\int \frac{d^3q}{(2\pi\hbar)^3} 
\matrixel{ps,\,qs'}{G(\omega = \varepsilon (p,s) + \varepsilon (q,s')}
{ps,\,qs'}_A n(q,s') \,,\label{eq:epsbhf}
\end{equation}
where $G$-matrix is the solution of the Bethe-Goldstone equation for a
realistic model of the NN interaction, $\omega$ is the starting energy  and
$n(q,s')$ is the occupation probability of nucleons with momentum $q$ and spin
projection $s'$. In the BHF approximation for systems at temperature $T=0$ this 
occupation probability is represented by a step function which restricts the
occupation to states with momenta below the Fermi momentum $k_F(s')$ of nucleons
with spin projection $s'$. Note that in order to shorten the notation, we have 
suppressed all labels referring to the isospin of the nucleons.

It is well known that a realistic NN interaction contains strong tensor and
spin-orbit components, which couple the spin and momentum 
or spatial dependence of the interaction. 
In the case of the deuteron these tensor components lead to
dumbbell shape if one considers deuterons with spin projection $M_s=\pm 1$
\cite{forest}. These considerations lead to the conjecture that also the
momentum distribution of polarized nuclear matter could be deformed with
respect to the axis of spin polarization. Configurations with deformed momentum
distributions could be lower in energy than the spherical ones discussed so
far, which would affect the polarizability. A deformed momentum distribution
would also be connected with the single-particle spectrum, in which the
single-particle energies depend on the magnitude and the direction of the
momentum $\vec p$ of the nucleon under consideration. This would lead to
different response functions and influence the mean free path of neutrinos as
well as other properties of the nuclear matter.

In order to investigate this question, we consider a Fermi momentum
that 
depends on the angle $\vartheta$ with respect to the spin-polarization axis,
parameterized in terms of
\begin{equation}
k_F(s,\vartheta ) = k_{F0}(s) \left(1 + \delta(s) \cos^2\vartheta \right)\,,
\label{eq:parakf}
\end{equation}
a parameterization which turned out to be useful. With this assumption we can
calculate the contribution of the single-particle potential $U(\vec p,s)$ to 
the single-particle energy in (\ref{eq:epsbhf}), employing the partial wave
expansion for the $G$-matrix as outlined e.g.~by Haftel and Tabakin\cite{haftel}
\begin{eqnarray}
U(\vec p,s) & = &\sum_{s',{\cal S},L',L,J,M_L,M} 
\int \frac{d^3q}{(2\pi\hbar)^3}\, n(\vec q,s')\, 
(L'M_L,{\cal S}s+s'\vert JM)  (LM_L,{\cal S}s+s'\vert JM)\nonumber\\
&& \qquad \times Y_{L'M_L}^* (\hat k_r)
Y_{LM_L} (\hat k_r) i^{(L'-L)} G_{L'L}^{J{\cal S}}(\omega, K_{CM}, k_r) 
\,,\label{eq:partwu}
\end{eqnarray}
with $\vec K_{cm}$ and $\vec k_r$ referring to the center of mass and relative
momentum of the interacting pair of nucleons, respectively. The brackets 
$(Lm_L,{\cal S},M_S\vert JM)$ denote the Clebsch-Gordan coefficient for the
coupling of the spin and orbital angular momentum of the relative motion.  Note
that for a system without spin-polarization the spherical harmonics  $Y_{L,M}$
drop out. If, however, we assume a spin-polarization Eq. (\ref{eq:partwu})
yields a single-particle potential which depends not only on the magnitude of
the momentum $\vec p$ but also on the angle relative to the axis of
spin-polarization, even if we assume spherical momentum distributions $n(\vec
q,s')$, i.e.~$\delta(s')=0$ in (\ref{eq:parakf}). From the momentum dependence
of the resulting single-particle energies, one can find a Fermi momentum
according to (\ref{eq:parakf}) and recalculate the single-particle energy with
this deformed momentum distribution until a self-consistent solution  is
obtained.

\begin{figure}
\begin{center}
\epsfig{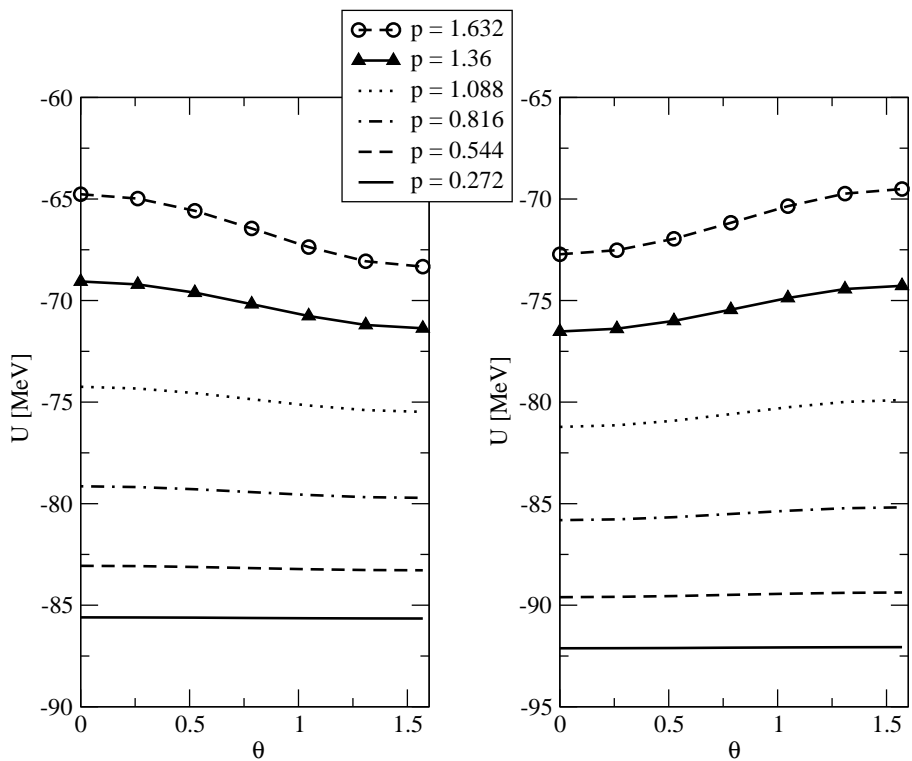}
\end{center}
\caption{Single-particle potentials for nucleons with s
pin parallel (left panel)
and antiparallel (right panel) to the spin polarization axis as a function of
the angle $\theta$ between their momentum and the polarization axis
in nuclear matter. Results are
given for various values for the modulus of the momentum $p$ (expressed in
fm$^{-1}$ as indicated in the legend). The single-particle potentials are
symmetric around $\theta = \pi/2$.\label{fig1}}
\end{figure}

Results for the single-particle potential of such BHF calculations are displayed
in Fig.~\ref{fig1} for nuclear matter at the empirical
saturation density $\rho_0$ with a spin polarization 
\begin{equation}
\Delta\rho = \frac{\rho^\uparrow - \rho^\downarrow}{\rho_0}
\end{equation} 
of 25 percent. The calculation have been performed using the charge-dependent 
CDBonn NN interaction of Machleidt et al.\cite{cdb}. The spin asymmetry as well
as the angle-dependence of the single-particle spectrum have been ignored in
the Pauli operator and starting energy, which were used to solve the
Bethe-Goldstone equation. The single-particle potential for the nucleons with
spin parallel to the polarization axis (left panel of Fig.~\ref{fig1}) and a
momentum of 1.46 fm$^{-1}$ (the Fermi momentum of the spin up momentum
distribution if it would be spherical) is about 2.7 MeV more attractive in the
direction of the momentum perpendicular to the polarization axis ($\theta =
\pi/2$) than for a momentum parallel or antiparallel to this axis. This means
that we obtain an oblate momentum distribution for the nucleons with spin
parallel to the polarization axis, which can be parameterized according to
Eq. (\ref{eq:parakf}) with a deformation parameter $\delta(\uparrow) \approx
-0.03$. It is worth noting that this oblate deformation is in line with the
observation made for the deuteron\cite{forest} as the dumbbell shape of the
density distribution mentioned above corresponds to a momentum distribution
with oblate deformation. 

The situation is opposite for the nucleons with spin antiparallel to the
polarization axis. In this case we observe a smaller single-particle energy for
momenta parallel to the polarization axis as compared to the momentum
perpendicular to this axis (see right panel of Fig.~\ref{fig1}). This leads to
a prolate deformation of the momentum distribution of the nucleon with spin
antiparallel to the polarization axis. 

The angle-dependence of the single-particle potential is of course much stronger
in the case of totally polarized nuclear matter. In this case we obtain a energy
difference of around 20 MeV between the single-particle energies for the 
direction of the momentum parallel as compared to perpendicular to the
polarization axis, if we consider a momentum of 1.72 fm$^{-1}$, which is the
Fermi momentum of the spherical distribution for this polarization. This
corresponds to a deformation parameter $\delta(\uparrow) \approx -0.15$ in the
parameterization of Eq. (\ref{eq:parakf}).  

The effects are weaker for pure neutron matter. For example,
in the case of completely polarized neutron matter at density $\rho_0$, the 
gain in the single-particle energy is around 6 MeV if the momentum is
perpendicular to the polarization axis $z$ as compared to the same momentum
parallel to $z$ (assuming a neutron with spin parallel to $z$ and a momentum 
$p=2.16$ fm$^{-1}$ which is the Fermi momentum of the spherical distribution).
This corresponds to a deformation parameter $\delta(\uparrow) \approx -0.03$.
    
To understand the origin of this deformation we recall that within the
Hartree-Fock approximation the non-locality or momentum dependence of the
self-energy of a nucleon in nuclear  matter is mainly due to the Fock-exchange
contribution in (\ref{eq:epsbhf}). For the sake of simplicity let us
consider completely polarized
nuclear matter. In this case the contribution of
the one-pion exchange to the Fock term in Eq. (\ref{eq:epsbhf}) for a nucleon
with spin parallel to the polarization axis can be described by
\begin{equation}
\Delta U_{\pi}(\vec p,\uparrow) \sim  \int d^3q \,\left\langle {\vec
p\uparrow,\,\vec q \uparrow}\Bigg| \frac {(\vec \sigma_1 \cdot \vec k)(\vec \sigma_2
\cdot \vec k)}{m_{\pi}^2 + k^2}\Bigg|  {\vec q\uparrow,\,\vec p\uparrow}
\right\rangle
n(\vec q,\uparrow)\,.\label{eq:pion1}
\end{equation}
In this equation we have approximated the one-pion exchange term by its static,
nonrelativistic reduction in which $\vec\sigma_i$ represents the vector of Pauli
matrices acting on nucleon $i$ and $\vec k$ denotes the momentum transfer which
is given in terms of the momenta of the interacting nucleons by
$\vec k = \vec p - \vec q$.
Since the spin part of the matrix element in (\ref{eq:pion1}) yields results
different from zero only for the $z$-component of the spin matrices ($z$-axis
parallel to the polarization axis) one obtains
\begin{equation}
\Delta U_{\pi}(\vec p,\uparrow) \sim \int d^3q \frac{\left(p_z - q_z\right)^2}
{m_{\pi}^2 + k^2}\,n(\vec q,\uparrow)\,.\label{eq:pion2}
\end{equation}
Indeed, we find that the self-energy for completely spin-polarized nuclear 
matter increases with the absolute value of $p_z$, but is rather weakly
dependent on $p_x$ and $p_y$. 

The situation is different, if we consider the self-energy of a nucleon with
spin antiparallel to the axis of spin polarization. In this case we have to
consider matrix elements of the form 
$$
\int d^3q \,\left\langle {\vec
p\downarrow,\,\vec q \uparrow}\Bigg| \frac {(\vec \sigma_1 \cdot \vec k)(\vec \sigma_2
\cdot \vec k)}{m_{\pi}^2 + k^2}\Bigg|  {\vec q\uparrow,\,\vec p\downarrow}
\right\rangle
n(\vec q,\uparrow)\, ,
$$ 
which lead to a self-energy which increases with the component of $\vec p$
perpendicular to the $z$-axis. This means that the angle-dependence of the
self-energy can be understood qualitatively in terms of the Fock contribution
of the pion-exchange part of the NN interaction. 

We will turn next to the effect of the deformed momentum
distribution on the total energy. The energy per nucleon is given by
\begin{equation}
\frac{E}{A} = \frac{\sum_s \int d^3p\, n(\vec p,s)
\frac{1}{2}\left(\frac{p^2}{2m} + \varepsilon(\vec p,s)\right)}
{\sum_s \int d^3p\, n(\vec p,s)}\,.\label{eq:EA}
\end{equation}
One finds that the gain in energy due to the deformation of the momentum
distribution is rather small (see table \ref{tab1}). Even in the case of
completely polarized nuclear matter, the deformed momentum distribution yields
an energy per nucleon which is 0.53 MeV smaller than the energy calculated for
the spherical momentum distribution. This is a rather small correction
compared to the energy difference between the polarized and unpolarized energy.

Therefore we cannot expect that the effects of deformed momentum distributions
will alter the conclusions of e.g.~\cite{vidana} about the equation of state for
spin polarized matter or the possible existence of a ferromagnetic state of
nuclear or neutron matter. The situation, however, might be compared to the
deformation degrees of freedom in finite nuclei. The possibility to form
deformed nuclei is not very relevant for the total energy of finite nuclei. The
gain in energy due to deformation of nuclei is typically of the order
of 0.05 MeV per nucleon or even less (see e.g.~\cite{MGAF}).  The deformation,
however, has rather significant effects on the excitation spectrum at low
energies, which is quite different for spherical as compared to deformed
nuclei. Also, the dependence of the single-particle energies on the direction of
the nucleon momentum in spin polarized nuclear matter may lead to quite
different response functions at low excitation energies. This may have a
significant effect e.g.~on the neutrino propagation in neutron stars with large
magnetic fields.

We like to acknowledge the financial support by the {\it Europ\"aische
Graduiertenkolleg T\"ubingen - Basel} (DFG - SNF) and the {\it
Sonderforschungsbereich 382} (DFG).

\begin{table}
\begin{center}
\begin{tabular}{c|r|rrr}
& \multicolumn{1}{c}{unpolarized} & \multicolumn{3}{c}{completely polarized} \\
& & \multicolumn{1}{c}{spherical} & \multicolumn{1}{c}{ non-sph.} & 
\multicolumn{1}{c}{$\delta(\uparrow)$}\\
\hline
Nuclear Matter & -16.46 MeV & 5.71 MeV & 5.17 MeV & -0.12 \\
Neutron Matter & 11.58 MeV & 56.94 MeV & 56.91 MeV & -0.03 \\
\end{tabular}
\caption{\label{tab1} Energy per nucleon for nuclear matter and neutron matter
of density $\rho_0$ = 0.17 fm$^{-3}$. Results are presented for the unpolarized
system and the system with complete spin polarization. For this case we show the
energies assuming a spherical and non-spherical momentum distribution. The last
column shows the deformation parameter for the self-consistent solution
according to Eq. (\protect\ref{eq:parakf})}
\end{center}
\end{table}
\vfil\eject
\end{document}